\documentclass[twoside]{article}
\usepackage{fleqn,espcrc2}
\usepackage{here,graphicx}

\newcommand{\figscale}{0.48}
\newcommand{\hatDS}{\hat{D}}
\newcommand{\Npoly}{N_{\mathit{poly}}}

\title{An exact algorithm for three-flavor QCD
       with $O(a)$-improved  Wilson fermions
       \thanks{Presented by K-I. Ishikawa}}

\newcommand{\Tsukuba}%
{Institute of Physics, University of Tsukuba, Tsukuba, Ibaraki 305-8571, Japan}

\newcommand{\RCCP}%
{Center for Computational Physics, University of Tsukuba, Tsukuba, Ibaraki 305-8577, Japan}

\newcommand{\ICRR}%
{Institute for Cosmic Ray Research, University of Tokyo, Kashiwa, Chiba 277-8582, Japan}

\newcommand{\KEK}%
{High Energy Accelerator Research Organization(KEK), Tsukuba, Ibaraki 305-0801, Japan}

\newcommand{\YITP}%
{Yukawa Institute for Theoretical Physics, Kyoto University, Kyoto 606-8502, Japan}
       
\author{
  JLQCD Collaboration: 
  S.~Aoki\address{\Tsukuba},
  R.~Burkhalter$^{\mathrm{a,}}$\address{\RCCP},
  M.~Fukugita\address{\ICRR},
  S.~Hashimoto\address{\KEK},
  K-I.~Ishikawa$^{\mathrm{d}}$,
  N.~Ishizuka$^{\mathrm{a,b}}$,
  Y.~Iwasaki$^{\mathrm{a,b}}$,
  K.~Kanaya$^{\mathrm{a,b}}$,
  T.~Kaneko$^{\mathrm{d}}$, 
  Y.~Kuramashi$^{\mathrm{d}}$,
  M.~Okawa$^{\mathrm{d}}$, 
  T.~Onogi\address{\YITP},
  S.~Tominaga$^{\mathrm{b}}$,
  N.~Tsutsui$^{\mathrm{d}}$,
  A.~Ukawa$^{\mathrm{a,b}}$,
  N.~Yamada$^{\mathrm{d}}$, 
  T.~Yoshi\'{e}$^{\mathrm{a,b}}$ }

\begin{document}

\begin{abstract}
We present an exact dynamical QCD simulation algorithm for
the $O(a)$-improved Wilson fermion with odd number of flavors.
Our algorithm is an extension of the non-Hermitian polynomials
HMC algorithm proposed by Takaishi and de Forcrand previously.
In our algorithm, the systematic errors caused by the polynomial
approximation of the inverse of Dirac operator is removed by 
a noisy-Metropolis test. For one flavor quark it is achieved by
taking the square root of the correction matrix explicitly.
We test our algorithm for the case of $N_f=1+1$ on a moderately large 
lattice size ($16^3\times48$). The $N_f=2+1$ case is also investigated.

{\large              
\vspace*{-30em}      
\hfill KEK-CP-112    
\vspace*{30em}       
}                    
\end{abstract}

\maketitle

\vspace*{-4em}
\section{Introduction}
\vspace*{-0.5em}
Lattice QCD simulations with three flavors of dynamical quarks are 
indispensable to understand the low energy QCD dynamics in the real world.
Takaishi and de Forcrand~\cite{Takaishi_Forcrand} have 
proposed a Hybrid Monte Carlo (HMC) algorithm which can treat
odd number of flavors of dynamical quarks, in which a non-Hermitian 
polynomial approximation is applied to the inverse of the Wilson-Dirac
operator.  Very recently they have removed the systematic error from
the polynomial approximation, making their algorithm 
exact~\cite{New_Takaishi_Forcrand}.

For realistic simulations with the available computational power,
the $O(a)$-improvement program is widely advocated.
We then extend the algorithm of 
Refs.~\cite{Takaishi_Forcrand,New_Takaishi_Forcrand} 
to the $O(a)$-improved Wilson quark actions.
We also develop a noisy-Metropolis test to remove the systematic error 
from the polynomial approximation, which is different from that of 
Ref.~\cite{New_Takaishi_Forcrand}.

In this article, we describe our algorithm for single flavor of dynamical 
quark. The consistency with the usual algorithm and applicability to 
large-scale simulations
are examined by running the $N_f$$=$$1$$+$$1$ and $N_f$$=$$2$ simulations 
on a $16^3$$\times$$48$ lattice. 
Employing the algorithm, we have started a search for the lattice parameters 
($\beta, \kappa$ etc.) suitable for realistic simulations with 
$N_f$$=$$2$$+$$1$ flavors of quarks (\textit{i.e.,} degenerate 
up-down quarks and strange quark).  We briefly describe the novel 
findings from the search, referring to a separate report~\cite{OKAWA} 
for details. 

\vspace*{-0.5em}
\section{Algorithm}
\label{sec:II}
\vspace*{-0.5em}
Our HMC algorithm for single flavor of dynamical quark 
is based on the following QCD partition function:
\begin{equation}
{\cal Z}=\int{\cal D}[U]
\det[(1+T)]\det[\hatDS_{oo}]
e^{-S_{\mathrm{g}}[U]},
\label{eq:PartitionFunc}
\end{equation}
where $S_{\mathrm{g}}[U]$ is the plaquette 
gauge action
and $\hat{D}_{oo}$ is the symmetrically preconditioned 
$O(a)$-improved Wilson-Dirac operator defined by
\begin{equation}
\hatDS_{oo}= 1-(1+T)^{-1}_{oo}M_{oe}(1+T)^{-1}_{ee} M_{eo},
\end{equation}
where the subscript $e$ ($o$) means the even (odd) 
components of the Dirac operator. The factor
$T$ is the local SW-term of the $O(a)$-improved Wilson-Dirac 
operator and $M$ is the hopping matrix.
There is another even-odd preconditioning in which only
the even-even (or odd-odd) SW-term is asymmetrically factored out.
We compared the efficiency of 
the symmetric and asymmetric preconditionings within the usual 
$N_f$$=$$2$ HMC algorithm, and found that the 
former has better efficiency. 
Therefore we employ the symmetrically
preconditioned form for the QCD partition function.
(For other preconditioning techniques, see Ref.~\cite{Peardon}.)

Using the trick described in 
Refs.~\cite{Takaishi_Forcrand,New_Takaishi_Forcrand,Alexandrou},
we rewrite the determinant of the Dirac operator as
\begin{equation}
\det[\hatDS_{oo}]\!=\!
\frac{\scriptstyle\det[\hatDS_{oo} P_{\Npoly}[\hatDS_{oo}]]}
     {\scriptstyle\det[P_{\Npoly}[\hatDS_{oo}]]}
\!=\!
\frac{\scriptstyle\det[W_{oo}]}
     {\scriptstyle|\det[T_{\Npoly}[\hatDS_{oo}]]|^{2}},
\label{eq:DetT}
\end{equation}
where $P_{\Npoly}[z]$ is a non-Hermitian polynomial of order 
$\Npoly$ (assumed to be even)
defined by $P_{\Npoly}[z]$$=$$\sum_{i=0}^{\Npoly}$$c_{i}(z\!-\!1)^{i}$ 
which approximates $1/z$, and
$W_{oo}$$\equiv$$\hatDS_{oo}$$P_{\Npoly}$$[\hatDS_{oo}]$.
$T_{\Npoly}[z]$ is a kind of square root of $P_{\Npoly}[z]$ 
defined by $P_{\Npoly}[z]$$=$ $T^{*}_{\Npoly}[z]$$T_{\Npoly}[z]$ with
$T_{\Npoly}[z]$$=$$\sum_{i=0}^{\Npoly/2}$ $d_{i}(z\!-\!1)^{i}$.
We employ the Chebyshev polynomial approximation
(\textit{i.e.,} the hopping matrix expansion)
for $P_{\Npoly}[z]$ and hence the coefficients 
are $c_{i}$$=$$(-1)^{i}$.
We use the Clenshaw's recurrence formula to calculate 
these polynomials.

Introducing pseudo fermion fields $\psi_{o}$ for 
$|\det[T_{\Npoly}[\hatDS_{oo}]]|^{2}$ in Eq.~(\ref{eq:DetT}),
Eq.~(\ref{eq:PartitionFunc}) becomes
\begin{eqnarray}
{\cal Z}&=&\int{\cal D}[U,\psi_{o}]
\det[W_{oo}]
e^{-S_{\mathit{eff}[U,\psi_{o}]}},\nonumber\\
S_{\mathit{eff}}[U,\psi_{o}]&=&S_{\mathrm{g}}[U]+
S_{\mathrm{det}}[U]+S_{\mathrm q}[U,\psi_{o}],\nonumber\\
S_{\mathrm{det}}[U]&=&-\log[\det[(1+T)]],\nonumber\\
S_{\mathrm{q}}[U,\psi_{o}]&=&|T_{\Npoly}[\hatDS_{oo}]\psi_{o}|^{2}.
\label{eq:PartitionPoly}
\end{eqnarray}
Our PHMC algorithm consists of the following two steps in this case:
1) the usual HMC algorithm for the effective
action $S_{\mathit{eff}}$,
2) the noisy-Metropolis test for the correction term $\det[W_{oo}]$.
The noisy-Metropolis test is carried out only after accepting the HMC 
Metropolis test.

The acceptance  probability of the noisy-Metropolis test is defined by
\begin{eqnarray}
&&P_{\mathit{corr}}[U\rightarrow U']=\min[1,e^{-dS}],\nonumber\\
&&dS=| A_{oo}[U']^{-1}A_{oo}[U] \chi_{o}|^2-|\chi_{o}|^2,
\label{eq:dS}
\end{eqnarray}
where $U$ is an initial configuration and $U'$ is a trial configuration
generated by the preceding HMC algorithm. 
Here $\chi_{o}$ is a random vector with Gaussian distribution, and 
$A_{oo}$ is defined by $A_{oo}^2$$=$$W_{oo}$.
This algorithm estimates $|\det[A_{oo}]|^{2}$ instead of $\det[W_{oo}]$.
$A_{oo}$ and $A_{oo}^{-1}$ are evaluated by the Taylor expansion with
respect to $W_{oo}-1$. 
In order to keep the exactness of our algorithm 
the residual of the Taylor expansion 
is monitored whenever Eq.~(\ref{eq:dS}) is calculated.
Our programs are written in double precision arithmetic, 
and optimized for HITACHI SR8000 at KEK.

\vspace*{-0.5em}
\section{$N_f=1+1$}
\label{sec:III}
\vspace*{-0.5em}
We test our single-flavor algorithm by additively combining 
two single-flavored pseudo fermions ($N_f$$=$$1$$+$$1$) and comparing 
results with those of the usual $N_f$$=$$2$ HMC algorithm.
In this case our algorithm runs as follows:
1) the HMC algorithm with the two single-flavored pseudo fermions, 2) 
the noisy-Metropolis test for each correction terms, which is carried out
only when the preceding Metropolis test is accepted.
To check viability of our algorithm toward realistic simulations, 
we employ a moderately large-scale simulation parameter:
$16^3$$\times$$48$, $\beta$$=$$5.2$, 
$c_{\mathrm{SW}}$$=$$2.02$, $\kappa$$=$$0.1340$ and $\kappa$$=$$0.1350$. 
These hopping parameters correspond to $m_{\pi}/m_{\rho}\sim$$0.8$ and 
$\sim$$0.7$, respectively.

\begin{figure}[t]
\centering
\includegraphics[scale=\figscale,clip]{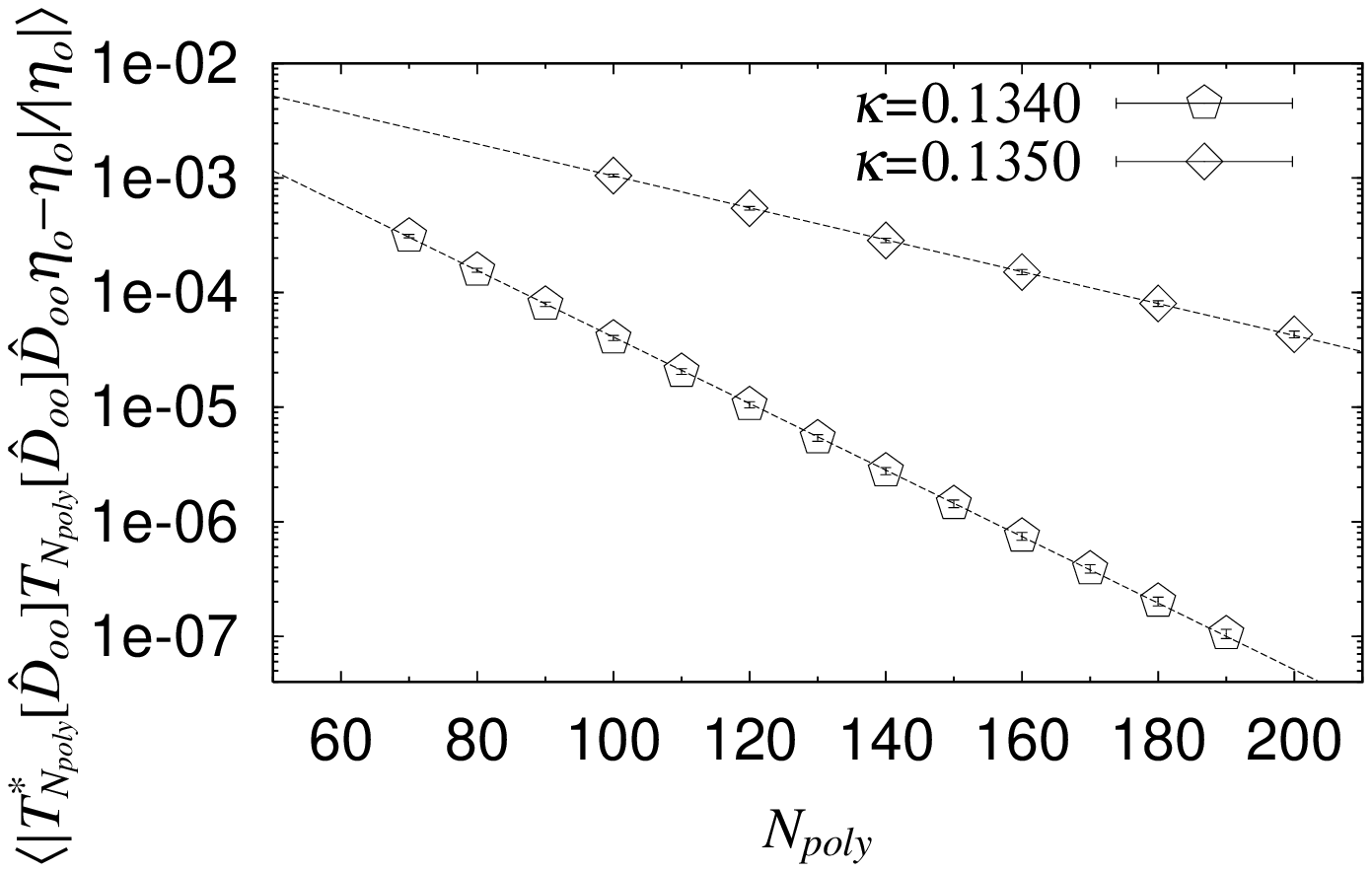}
\vspace*{-3em}
\caption{Convergence $T_{\Npoly}[\hatDS_{oo}]$ as $\Npoly\to\infty$. }
\label{fig:1}
\vspace*{-3em}
\end{figure}

Figure~\ref{fig:1} shows the convergence of $T_{\Npoly}[\hatDS_{oo}]$
as $\Npoly$ increases.  
The measurement is made on 20 configurations separated by 10 trajectories.
The residual is defined by 
$|T^{*}_{\Npoly}[\hatDS_{oo}]$$T_{\Npoly}[\hatDS_{oo}]$$\hatDS_{oo}$$\eta_{o}$$-$$\eta_{o}|$$/$$|\eta_{o}|$
with a Gaussian noise vector $\eta_{o}$.
We observe an exponential decay as expected and there are no accumulation of
round-off errors for large $\Npoly$. 

\begin{figure}[t]
\centering
\includegraphics[scale=\figscale,clip]{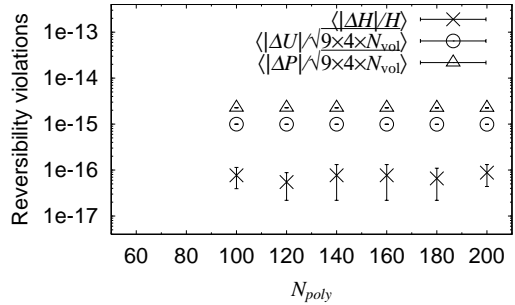}
\vspace*{-3em}
  \caption{$\Npoly$ dependence of the reversibility violations 
  at $\kappa=0.1350$.
  $|\Delta H|/H$ : total Hamiltonian, $\Delta U$: gauge link, $\Delta P$: gauge momentum.}
  \label{fig:2}
\vspace*{-2em}
\end{figure}

We also investigate the reversibility of the molecular dynamics step and 
observe that the violation stays around the limit of double precision
for both of the quark masses 
(see Fig.~\ref{fig:2} for $\kappa$$=$ $0.1350$). 
The value of plaquette, averaged over $\sim$$1000$ 
trajectories, is consistent between the two algorithms 
(\textit{e.g.}, $\langle P\rangle$$=$$0.53393(11)$(HMC) and  
                                  $0.53392(9)$(PHMC with $\Npoly$$=$$140$)
for $\kappa$$=$$0.1350$).
With these observation we conclude that
our PHMC algorithm for single-flavor dynamical quark works well on moderately
large lattices and intermediate quark mass 
for at least $N_f$$=$$1$$+$$1$. 


\vspace*{-0.5em}
\section{$N_f=2+1$}
\label{sec:IV}
\vspace*{-0.5em}
For the $N_f$$=$$2$$+$$1$ case, we additively combine 
the standard two-flavored 
pseudo fermion and our single-flavored pseudo fermion as suggested in 
Refs.~\cite{Takaishi_Forcrand,New_Takaishi_Forcrand}.
The algorithm is given by 1) the HMC step with the combined effective 
Hamiltonian, and 2) the noisy-Metropolis test for the correction factor.

\begin{figure}[t]
\centering
\includegraphics[scale=\figscale,clip]{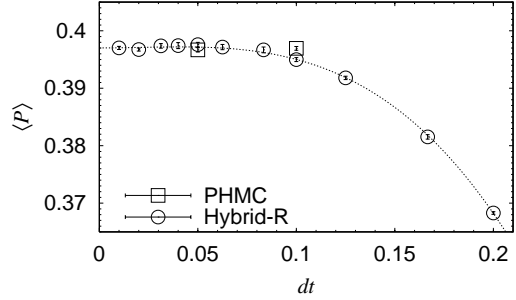}
\vspace*{-3em}
\caption{Plaquette vs $dt$.}
\vspace*{-2em}
\label{fig:3}
\end{figure}

We compare results for the averaged plaquette 
between our algorithm and the Hybrid-R algorithm
on a $4^3$$\times$$8$ lattice at $\beta$$=$$4.8$,
$c_{\mathrm{SW}}$$=$$1.0$, $\kappa_{ud}$$=$$0.150$, and 
$\kappa_s$$=$$0.140$.
Figure~\ref{fig:3} shows the molecular dynamics step size $dt$ dependence
of the plaquette value with the Hybrid-R and PHMC($\Npoly$$=$$10$) algorithms.
The results with the PHMC algorithm (squares) 
do not depend on $dt$ and are consistent
with the value at $dt\to 0$ with the Hybrid-R algorithm
($\langle P\rangle$$=$$0.39669(38)$(PHMC), $0.39702(12)$(Hybrid-R)).

Encouraged with these results, we haved performed a series of parameter 
searches that would realize $a^{-1}$$\sim$$1$--$2$ GeV, $L$$\sim$$1$--$2$ fm, 
$m_{\pi}/m_{\rho}$$\sim$$0.7$--$0.8$
on a $12^3$$\times$$32$ lattice.  
During the search, which employed the tadpole-improved one-loop 
value for $c_{\mathrm{SW}}$ and degenerate quark mass ($N_f$$=$$3$), 
we have found an unexpected first-order phase transition around 
$\beta$$\sim$$4.8$--$5.0$ . 
This phase transition is also observed with the Hybrid-R algorithm 
on a $8^3$$\times$$16$ lattice with the same lattice parameter.
The details of the phase structure are given in Ref.~\cite{OKAWA}.

\vspace*{3mm}
This work is supported by the Supercomputer Project No.66 (FY2001) of High
Energy Accelerator Research Organization (KEK), and also in part by the
Grant-in-Aid of the Ministry of Education (Nos. 10640246, 11640294, 12014202,
12640253, 12640279, 12740133, 13640260 and 13740169). K-I.I. and N.Y. are
supported by the JSPS Research Fellowship.

\end{document}